\PassOptionsToPackage{unicode}{hyperref}
\PassOptionsToPackage{hyphens}{url}
\PassOptionsToPackage{dvipsnames,svgnames,x11names}{xcolor}
\documentclass[
]{article}
\usepackage{amsmath,amssymb}
\usepackage{lmodern}
\usepackage{iftex}
\ifPDFTeX
  \usepackage[T1]{fontenc}
  \usepackage[utf8]{inputenc}
  \usepackage{textcomp} 
\else 
  \usepackage{unicode-math}
  \defaultfontfeatures{Scale=MatchLowercase}
  \defaultfontfeatures[\rmfamily]{Ligatures=TeX,Scale=1}
\fi
\IfFileExists{upquote.sty}{\usepackage{upquote}}{}
\IfFileExists{microtype.sty}{
  \usepackage[]{microtype}
  \UseMicrotypeSet[protrusion]{basicmath} 
}{}
\makeatletter
\@ifundefined{KOMAClassName}{
  \IfFileExists{parskip.sty}{%
    \usepackage{parskip}
  }{
    \setlength{\parindent}{0pt}
    \setlength{\parskip}{6pt plus 2pt minus 1pt}}
}{
  \KOMAoptions{parskip=half}}
\makeatother
\usepackage{xcolor}
\usepackage[margin=1in]{geometry}
\usepackage{color}
\usepackage{fancyvrb}

\DefineVerbatimEnvironment{Highlighting}{Verbatim}{commandchars=\\\{\}}
\usepackage{framed}
\definecolor{shadecolor}{RGB}{248,248,248}
\newenvironment{Shaded}{\begin{snugshade}}{\end{snugshade}}

\newcommand{\AttributeTok}[1]{\textcolor[rgb]{0.77,0.63,0.00}{#1}}

\newcommand{\CommentTok}[1]{\textcolor[rgb]{0.56,0.35,0.01}{\textit{#1}}}

\newcommand{\ConstantTok}[1]{\textcolor[rgb]{0.00,0.00,0.00}{#1}}
\newcommand{\ControlFlowTok}[1]{\textcolor[rgb]{0.13,0.29,0.53}{\textbf{#1}}}

\newcommand{\DecValTok}[1]{\textcolor[rgb]{0.00,0.00,0.81}{#1}}

\newcommand{\FloatTok}[1]{\textcolor[rgb]{0.00,0.00,0.81}{#1}}
\newcommand{\FunctionTok}[1]{\textcolor[rgb]{0.00,0.00,0.00}{#1}}

\newcommand{\NormalTok}[1]{#1}

\newcommand{\OtherTok}[1]{\textcolor[rgb]{0.56,0.35,0.01}{#1}}

\newcommand{\SpecialCharTok}[1]{\textcolor[rgb]{0.00,0.00,0.00}{#1}}

\usepackage{longtable,booktabs,array}
\usepackage{calc} 
\usepackage{etoolbox}
\makeatletter
\patchcmd\longtable{\par}{\if@noskipsec\mbox{}\fi\par}{}{}
\makeatother
\IfFileExists{footnotehyper.sty}{\usepackage{footnotehyper}}{\usepackage{footnote}}
\makesavenoteenv{longtable}
\usepackage{graphicx}
\makeatletter
\def\maxwidth{\ifdim\Gin@nat@width>\linewidth\linewidth\else\Gin@nat@width\fi}
\def\maxheight{\ifdim\Gin@nat@height>\textheight\textheight\else\Gin@nat@height\fi}
\makeatother
\setkeys{Gin}{width=\maxwidth,height=\maxheight,keepaspectratio}
\makeatletter
\def\fps@figure{htbp}
\makeatother
\providecommand{\tightlist}{%
  \setlength{\itemsep}{0pt}\setlength{\parskip}{0pt}}
\ifLuaTeX
\usepackage[bidi=basic]{babel}
\else
\usepackage[bidi=default]{babel}
\fi
\babelprovide[main,import]{american}

\def\languageshorthands#1{}
\usepackage{hyperref}
\definecolor{linkcolor}{HTML}{D55E00}
\definecolor{citecolor}{HTML}{009E73}
\definecolor{urlcolor}{HTML}{0072B2}
\hypersetup{
    colorlinks,
    linkcolor={linkcolor},
    citecolor={citecolor},
    urlcolor={urlcolor}
}

\makeatletter
\@ifundefined{DecValTok}{}{
  \renewcommand{\DecValTok}[1]{\textcolor[HTML]{009E73}{#1}}
  \renewcommand{\FloatTok}[1]{\textcolor[HTML]{009E73}{#1}}
  \renewcommand{\ConstantTok}[1]{\textcolor[HTML]{009E73}{#1}}
  \renewcommand{\ControlFlowTok}[1]{\textcolor[HTML]{0072B2}{\textbf{#1}}}
  \renewcommand{\OtherTok}[1]{\textcolor[HTML]{000000}{#1}}
  \renewcommand{\CommentTok}[1]{\textcolor[HTML]{999999}{\textit{#1}}}
  \renewcommand{\AttributeTok}[1]{\textcolor[HTML]{CC79A7}{#1}}
  \renewcommand{\FunctionTok}[1]{\textcolor[HTML]{56B4E9}{#1}}
}
\makeatother

\usepackage{caption}
\usepackage[absolute]{textpos}
\usepackage{amsmath}
\usepackage{float}
\usepackage{csquotes}
\usepackage{booktabs}
\usepackage{longtable}
\usepackage{array}
\usepackage{multirow}
\usepackage{wrapfig}
\usepackage{colortbl}
\usepackage{pdflscape}
\usepackage{tabu}
\usepackage{threeparttable}
\usepackage{threeparttablex}
\usepackage[normalem]{ulem}
\usepackage{makecell}
\usepackage{xcolor}
\ifLuaTeX
  \usepackage{selnolig}  
\fi
\usepackage[style=alphabetic,sorting=anyt]{biblatex}
\IfFileExists{bookmark.sty}{\usepackage{bookmark}}{\usepackage{hyperref}}
\IfFileExists{xurl.sty}{\usepackage{xurl}}{} 
\hypersetup{
  pdftitle={Trimmed Harrell-Davis quantile estimator based on the highest density interval of the given width},
  pdfauthor={Andrey Akinshin},
  pdflang={en-US},
  colorlinks=true,
  linkcolor={linkcolor},
  filecolor={Maroon},
  citecolor={citecolor},
  urlcolor={urlcolor},
  pdfcreator={LaTeX via pandoc}}

\title{Trimmed Harrell-Davis quantile estimator based on the highest density interval of the given width}
\author{Andrey Akinshin\\
JetBrains, \href{mailto:andrey.akinshin@gmail.com}{\nolinkurl{andrey.akinshin@gmail.com}}}
\date{}

\begin{document}
\maketitle
\begin{abstract}
Traditional quantile estimators that are based on one or two order statistics are a common way to estimate
distribution quantiles based on the given samples.
These estimators are robust, but their statistical efficiency is not always good enough.
A more efficient alternative is the Harrell-Davis quantile estimator which uses
a weighted sum of all order statistics.
Whereas this approach provides more accurate estimations for the light-tailed distributions, it's not robust.
To be able to customize the trade-off between statistical efficiency and robustness,
we could consider \emph{a trimmed modification of the Harrell-Davis quantile estimator}.
In this approach, we discard order statistics with low weights according to
the highest density interval of the beta distribution.

\textbf{Keywords:} quantile estimation, robust statistics, Harrell-Davis quantile estimator.
\end{abstract}

\begin{textblock*}{600mm}(.5\textwidth,1cm)
This is an original manuscript of an article published by Taylor \& Francis\\
in Communications in Statistics --- Simulation and Computation on 17 March 2022,\\
available online: \url{https://www.tandfonline.com/10.1080/03610918.2022.2050396}
\end{textblock*}

\hypertarget{introduction}{%
\section{Introduction}\label{introduction}}

We consider a problem of quantile estimation for the given sample.
Let \(x\) be a sample with \(n\) elements: \(x = \{ x_1, x_2, \ldots, x_n \}\).
We assume that all sample elements are sorted (\(x_1 \leq x_2 \leq \ldots \leq x_n\)) so that
we could treat the \(i^\textrm{th}\) element \(x_i\) as the \(i^\textrm{th}\) order statistic \(x_{(i)}\).
Based on the given sample, we want to build an estimation of the \(p^\textrm{th}\) quantile \(Q(p)\).

The traditional way to do this is to use a single order statistic
or a linear combination of two subsequent order statistics.
This approach could be implemented in various ways.
A classification of the most popular implementations could be found in \autocite{hyndman1996}.
In this paper, Rob J. Hyndman and Yanan Fan describe nine types of traditional quantile estimators
which are used in statistical computer packages.
The most popular approach in this taxonomy is Type 7 which is used by default in R, Julia, NumPy, and Excel:

\[
Q_{\operatorname{HF7}}(p) = x_{\lfloor h \rfloor}+(h-\lfloor h \rfloor)(x_{\lceil h \rceil}-x_{\lfloor h \rfloor}),
\quad h = (n-1)p+1.
\]

\begin{figure}[ht!]

{\centering \includegraphics{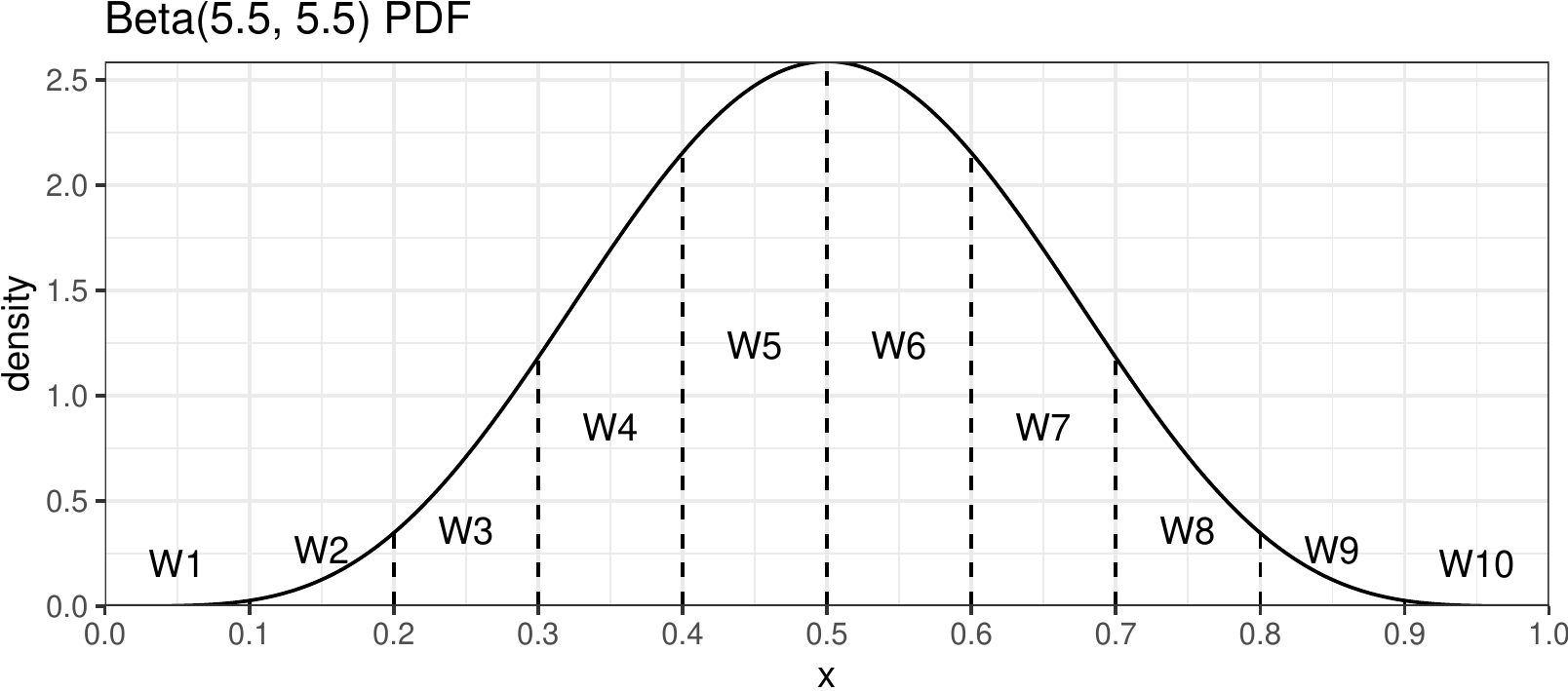} 

}

\caption{Beta distribution B(5.5, 5.5) probability density function}\label{fig:beta}
\end{figure}

Traditional quantile estimators have simple implementations and a good robustness level.
However, their statistical efficiency is not always good enough:
the obtained estimations could noticeably differ from the true distribution quantile values.
The gap between the estimated and true values could be decreased by increasing the number of used order statistics.
In \autocite{harrell1982}, Frank E. Harrell and C. E. Davis suggest estimating quantiles using
a weighted sum of all order statistics:

\[
Q_{\operatorname{HD}}(p) = \sum_{i=1}^{n} W_{\operatorname{HD},i} \cdot x_i,\quad
W_{\operatorname{HD},i} = I_{i/n}(\alpha, \beta) - I_{(i-1)/n}(\alpha, \beta),
\]

where \(I_x(\alpha, \beta)\) is the regularized incomplete beta function,
\(\alpha = (n+1)p\), \(\;\beta = (n+1)(1-p)\).
To get a better understanding of this approach,
we could look at the probability density function of the beta distribution \(\operatorname{Beta}(\alpha, \beta)\) (see Figure \ref{fig:beta}).
If we split the \([0;1]\) interval into \(n\) segments of equal width,
we can define \(W_{\operatorname{HD},i}\) as the area under curve in the \(i^\textrm{th}\) segment.
Since \(I_x(\alpha, \beta)\) is the cumulative distribution function of \(\operatorname{Beta}(\alpha, \beta)\),
we can express \(W_{\operatorname{HD},i}\) as \(I_{i/n}(\alpha, \beta) - I_{(i-1)/n}(\alpha, \beta)\).

The Harrell-Davis quantile estimator is suggested in \autocite{david2003}, \autocite{grissom2005}, \autocite{wilcox2016}, and \autocite{gibbons2020}
as an efficient alternative to the traditional estimators.
In \autocite{yoshizawa1985}, asymptotic equivalence for the traditional sample median and the Harrell--Davis median estimator is shown.

The Harrell-Davis quantile estimator shows decent statistical efficiency in the case of light-tailed distributions:
its estimations are much more accurate than estimations of the traditional quantile estimators.
However, the improved efficiency has a price: \(Q_{\operatorname{HD}}\) is not robust.
Since the estimation is a weighted sum of all order statistics with positive weights,
a single corrupted element may spoil all the quantile estimations, including the median.
It may become a severe drawback in the case of heavy-tailed distributions in which
it's a typical situation when we have a few extremely large outliers.
In such cases, we use the median instead of the mean as a measure of central tendency
because of its robustness.
Indeed, if we estimate the median using the traditional quantile estimators like \(Q_{\operatorname{HF7}}\),
its asymptotic breakdown point is 0.5.
Unfortunately, if we switch to \(Q_{\operatorname{HD}}\), the breakdown point becomes zero
so that we completely lose the median robustness.

Another severe drawback of \(Q_{\operatorname{HD}}\) is its computational complexity.
If we have a sorted array of numbers,
a traditional quantile estimation could be computed using \(O(1)\) simple operations.
For an unsorted sample,
there are approaches that allow getting an order statistic
using \(O(n)\) operations (e.g., see \autocite{alexandrescu2017}).
If we estimate the quantiles using \(Q_{\operatorname{HD}}\), we need \(O(n)\) operations for a sorted array.
Moreover, these operations involve computation of \(I_x(\alpha, \beta)\) values
which are pretty expensive from the computational point of view.

Alternatively, we could consider
the Sfakianakis-Verginis quantile estimator (see \autocite{sfakianakis2008}) or
the Navruz-Özdemir quantile estimator (see \autocite{navruz2020}) which
are also based on a weighted sum of all order statistics.
However, they also have the disadvantages listed above.

Neither \(Q_{\operatorname{HF7}}\) nor \(Q_{\operatorname{HD}}\) fit all kinds of problem.
\(Q_{\operatorname{HF7}}\) is simple, robust, and computationally fast,
but its statistical efficiency doesn't always satisfy the business requirements.
\(Q_{\operatorname{HD}}\) could provide better statistical efficiency,
but it's computationally slow and not robust.

To get a reasonable trade-off between \(Q_{\operatorname{HF7}}\) and \(Q_{\operatorname{HD}}\),
we consider a trimmed modification of the Harrell-Davis quantile estimator.
The core idea is simple:
we take the classic Harrell-Davis quantile estimator,
find the highest density interval of the underlying beta distribution,
discard all the order statistics outside the interval,
and calculate a weighted sum of the order statistics within the interval.
The obtained quantile estimation is more robust than \(Q_{\operatorname{HD}}\) (because it doesn't use extreme values)
and typically more statistically efficient than \(Q_{\operatorname{HF7}}\)
(because it uses more than only two order statistics).
Let's discuss this approach in detail.

\hypertarget{the-trimmed-harrell-davis-quantile-estimator}{%
\section{The trimmed Harrell-Davis quantile estimator}\label{the-trimmed-harrell-davis-quantile-estimator}}

The estimators based on one or two order statistics are not efficient enough because they use too few sample elements.
The estimators based on all order statistics are not robust enough because they use too many sample elements.
It looks reasonable to consider a quantile estimator based on a variable number of order statistics.
This number should be large enough to ensure decent statistical efficiency
but not too large to exclude possible extreme outliers.

A robust alternative to the mean is the trimmed mean.
The idea behind it is simple: we should discard some sample elements at both ends
and use only the middle order statistics.
With this approach, we can customize the trade-off between robustness and statistical efficiency
by controlling the number of the discarded elements.
If we apply the same idea to \(Q_{\operatorname{HD}}\),
we can build a trimmed modification of the Harrell-Davis quantile estimator.
Let's denote it as \(Q_{\operatorname{THD}}\).

\begin{figure}[ht!]

{\centering \includegraphics{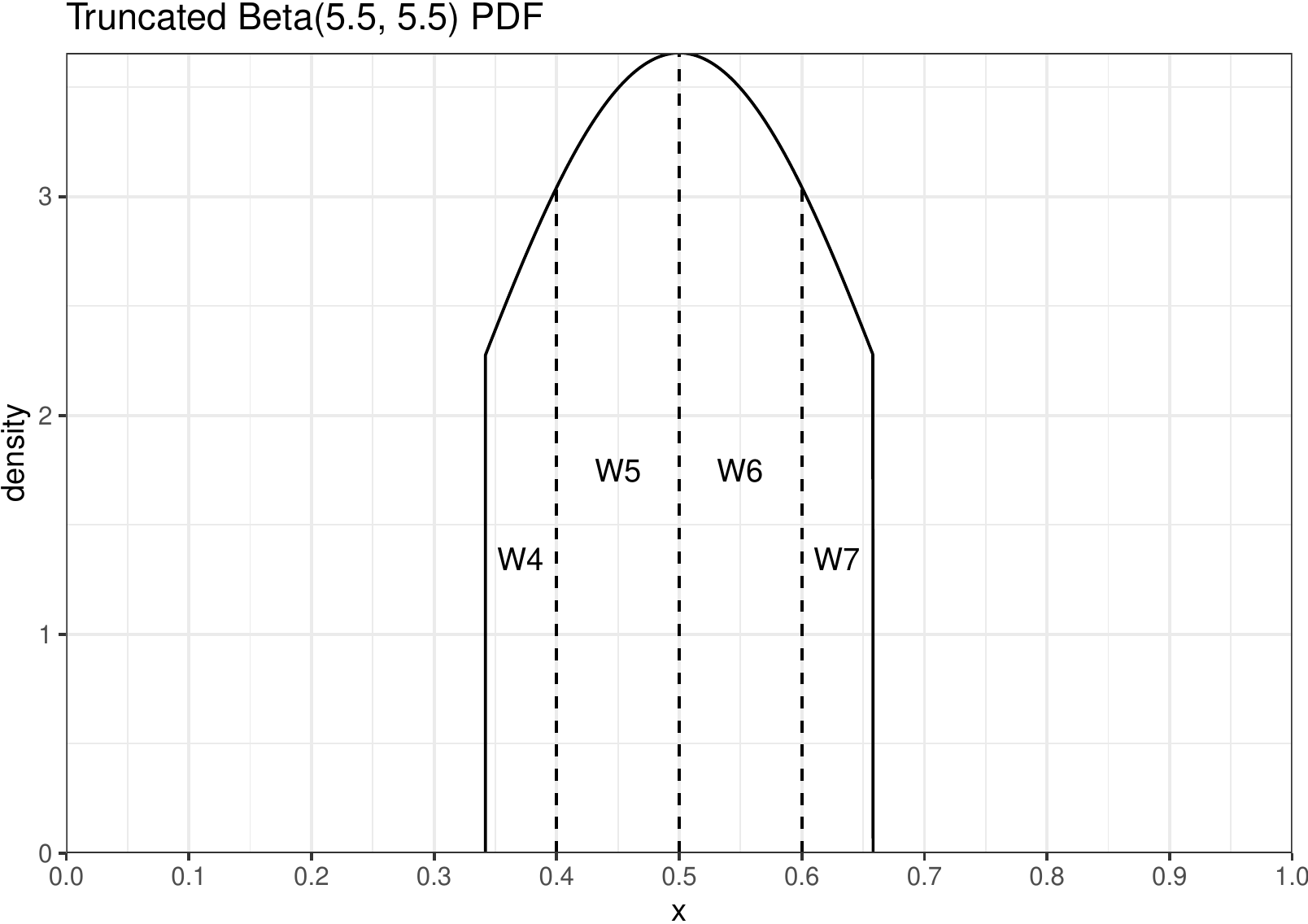} 

}

\caption{Truncated beta distribution B(5.5, 5.5) probability density function}\label{fig:tbeta}
\end{figure}

In the case of the trimmed mean, we typically discard the same number of elements on each side.
We can't do the same for \(Q_{\operatorname{THD}}\)
because the array of order statistic weights \(\{ W_{\operatorname{HD},i} \}\) is asymmetric.
It looks reasonable to drop the elements with the lowest weights and keep the elements with the highest weights.
Since the weights are assigned according to the beta distribution,
the range of order statistics with the highest weight concentration could be found
using the beta distribution highest density interval.
Thus, once we fix the proportion of dropped/kept elements,
we should find the highest density interval of the given width.
Let's denote the interval as \([L;R]\) where \(R-L=D\).
The order statistics weights for \(Q_{\operatorname{THD}}\) should be defined
using a part of the beta distribution within this interval.
It gives us the truncated beta distribution \(\operatorname{TBeta}(\alpha, \beta, L, R)\) (see Figure \ref{fig:tbeta}).

\clearpage

We know the CDF for \(\operatorname{Beta}(\alpha, \beta)\) which is used in \(Q_{\operatorname{HD}}\):
\(F_{\operatorname{HD}}(x) = I_x(\alpha, \beta)\).
For \(Q_{\operatorname{THD}}\),
we need the CDF for \(\operatorname{TBeta}(\alpha, \beta, L, R)\) which could be easily found:

\[
F_{\operatorname{THD}}(x) = \begin{cases}
0 & \textrm{for }\, x < L,\\
\big( F_{\operatorname{HD}}(x) - F_{\operatorname{HD}}(L) \big) /
\big( F_{\operatorname{HD}}(R) - F_{\operatorname{HD}}(L) \big)
  & \textrm{for }\, L \leq x \leq R,\\
1 & \textrm{for }\, R < x.
\end{cases}
\]

The final \(Q_{\operatorname{THD}}\) equation has the same form as \(Q_{\operatorname{HD}}\):

\[
Q_{\operatorname{THD}} = \sum_{i=1}^{n} W_{\operatorname{THD},i} \cdot x_i, \quad
W_{\operatorname{THD},i} = F_{\operatorname{THD}}(i / n) - F_{\operatorname{THD}}((i - 1) / n).
\]

There is only one thing left to do:
we should choose an appropriate width \(D\) of the beta distribution highest density interval.
In practical application, this value should be chosen based on the given problem:
researchers should \emph{carefully} analyze business requirements,
describe desired robustness level via setting the breakdown point,
and come up with a \(D\) value that satisfies the initial requirements.

However, if we have absolutely no information about the problem, the underlying distribution,
and the robustness requirements, we can use the following rule of thumb which gives the starting point:
\(D=1/\sqrt{n}\).
We denote \(Q_{\operatorname{THD}}\) with such a \(D\) value as \(Q_{\operatorname{THD-SQRT}}\).
In most cases, it gives an acceptable trade-off between the statistical efficiency and the robustness level.
Also, \(Q_{\operatorname{THD-SQRT}}\) has a practically reasonable computational complexity:
\(O(\sqrt{n})\) instead of \(O(n)\) for \(Q_{\operatorname{HD}}\).
For example, if \(n=10\,000\), we have to process only 100 sample elements and calculate 101 values of \(I_x(\alpha, \beta)\).

\textbf{An example}
Let's say we have the following sample:

\[
x = \{ -0.565, -0.106, -0.095, 0.363, 0.404, 0.633, 1.371, 1.512, 2.018, 100\,000 \}.
\]

Nine elements were randomly taken from the standard normal distribution \(\mathcal{N}(0, 1)\).
The last element \(x_{10}\) is an outlier.
The weight coefficient for \(Q_{\operatorname{HD}}\) an \(Q_{\operatorname{THD-SQRT}}\)
are presented in Table \ref{tab:t1}.

\begin{longtable}[]{@{}
  >{\raggedleft\arraybackslash}p{(\columnwidth - 6\tabcolsep) * \real{0.0617}}
  >{\raggedleft\arraybackslash}p{(\columnwidth - 6\tabcolsep) * \real{0.1975}}
  >{\raggedleft\arraybackslash}p{(\columnwidth - 6\tabcolsep) * \real{0.3333}}
  >{\raggedleft\arraybackslash}p{(\columnwidth - 6\tabcolsep) * \real{0.4074}}@{}}
\caption{\label{tab:t1} Weight coefficients of \(Q_{\operatorname{HD}}\) and \(Q_{\operatorname{THD-SQRT}}\) for \(n=10\)}\tabularnewline
\toprule()
\begin{minipage}[b]{\linewidth}\raggedleft
\(i\)
\end{minipage} & \begin{minipage}[b]{\linewidth}\raggedleft
\(x_i\)
\end{minipage} & \begin{minipage}[b]{\linewidth}\raggedleft
\(W_{\operatorname{HD},i}\)
\end{minipage} & \begin{minipage}[b]{\linewidth}\raggedleft
\(W_{\operatorname{THD-SQRT},i}\)
\end{minipage} \\
\midrule()
\endfirsthead
\toprule()
\begin{minipage}[b]{\linewidth}\raggedleft
\(i\)
\end{minipage} & \begin{minipage}[b]{\linewidth}\raggedleft
\(x_i\)
\end{minipage} & \begin{minipage}[b]{\linewidth}\raggedleft
\(W_{\operatorname{HD},i}\)
\end{minipage} & \begin{minipage}[b]{\linewidth}\raggedleft
\(W_{\operatorname{THD-SQRT},i}\)
\end{minipage} \\
\midrule()
\endhead
1 & -0.565 & 0.0005 & 0 \\
2 & -0.106 & 0.0146 & 0 \\
3 & -0.095 & 0.0727 & 0 \\
4 & 0.363 & 0.1684 & 0.1554 \\
5 & 0.404 & 0.2438 & 0.3446 \\
6 & 0.633 & 0.2438 & 0.3446 \\
7 & 1.371 & 0.1684 & 0.1554 \\
8 & 1.512 & 0.0727 & 0 \\
9 & 2.018 & 0.0146 & 0 \\
10 & \(100\,000.000\) & 0.0005 & 0 \\
\bottomrule()
\end{longtable}

Here are the corresponding quantile estimations:

\[
Q_{\operatorname{HD}}(0.5) \approx 51.9169, \quad Q_{\operatorname{THD}}(0.5) \approx 0.6268.
\]

As we can see, \(Q_{\operatorname{HD}}\) is heavily affected by the outlier \(x_{10}\).
Meanwhile, \(Q_{\operatorname{THD}}\) gives a reasonable median estimation
because it uses a weighted sum of four middle order statistics.

\hypertarget{beta-distribution-highest-density-interval-of-the-given-width}{%
\section{Beta distribution highest density interval of the given width}\label{beta-distribution-highest-density-interval-of-the-given-width}}

In order to build the truncated beta distribution for \(Q_{\operatorname{THD}}\),
we have to find the \(\operatorname{Beta}(\alpha, \beta)\) highest density interval of the required width \(D\).
Thus, for the given \(\alpha, \beta, D\), we should provide an interval \([L;R]\):

\[
\operatorname{BetaHDI}(\alpha, \beta, D) = [L; R].
\]

\begin{figure}[ht!]

{\centering \includegraphics{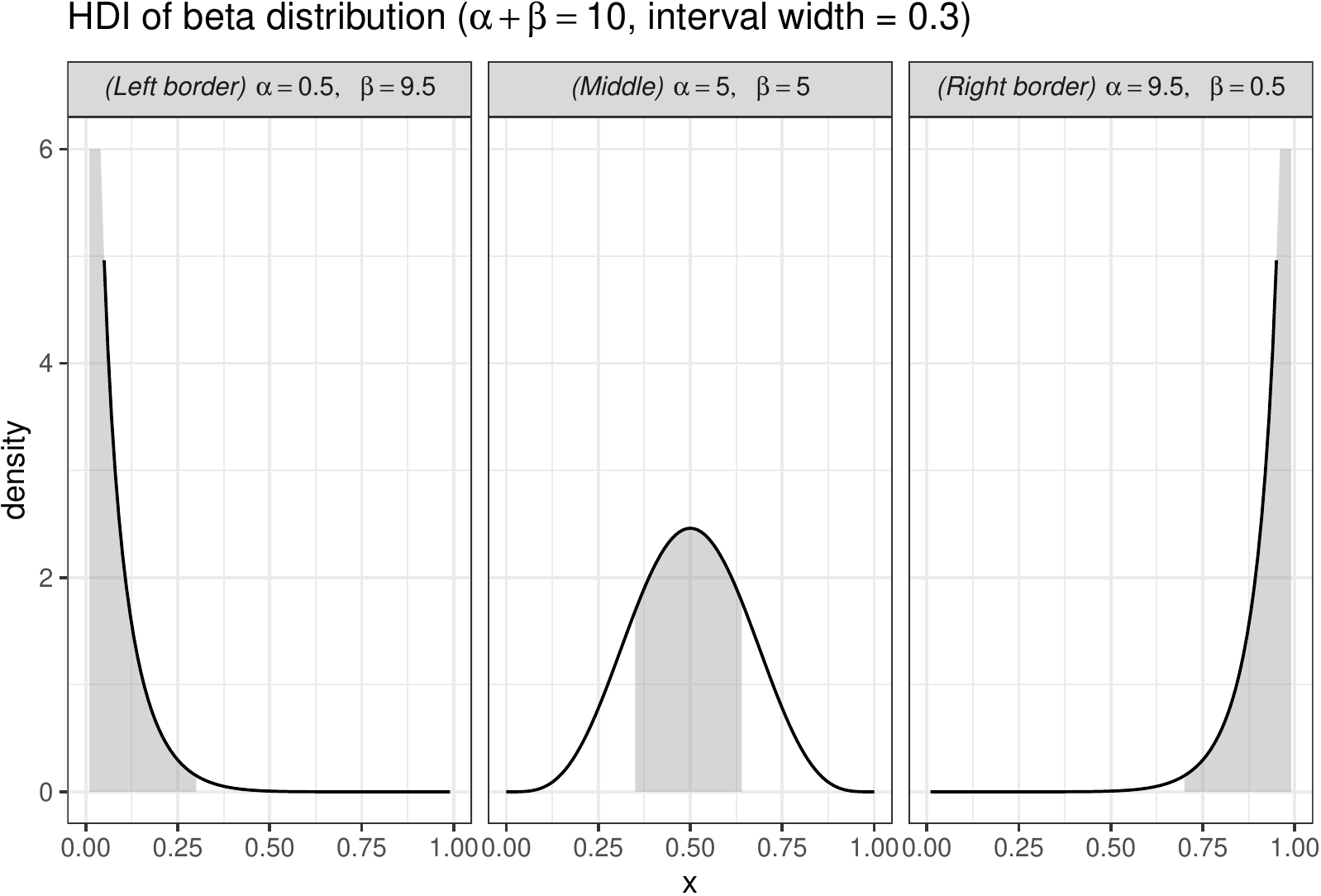} 

}

\caption{Highest density interval of the beta distribution based on the mode location}\label{fig:hdi}
\end{figure}

Let's briefly discuss how to do this.
First of all, we should calculate the mode \(M\) of \(\operatorname{Beta}(\alpha, \beta)\)
(non-degenarate cases are presented in Figure \ref{fig:hdi}):

\[
M = \operatorname{Mode}_{\alpha, \beta} =
\begin{cases}
  \{0, 1 \} \textrm{ or any value in } (0, 1) & \textrm{for }\, \alpha \leq 1,\, \beta \leq 1 \textit{ (Degenerate case),} \\
  0                                           & \textrm{for }\, \alpha \leq 1,\, \beta > 1    \textit{ (Left border case),} \\
  1                                           & \textrm{for }\, \alpha > 1,\, \beta \leq 1    \textit{ (Right border case),} \\
  \frac{\alpha - 1}{\alpha + \beta - 2}       & \textrm{for }\, \alpha > 1,\, \beta > 1       \textit{ (Middle case).}
\end{cases}
\]

\clearpage

The actual value of \(\operatorname{BetaHDI}(\alpha, \beta, D)\) depends on the specific case from the above list
which defines the mode location.
Three of these cases are easy to handle:

\begin{itemize}
\tightlist
\item
  Degenerate case \(\alpha \leq 1, \beta \leq 1\):
  There is only one way to get such a situation: \(n = 1, p = 0.5\).
  Since such a sample contains a single element, it doesn't matter how we choose the interval.
\item
  Left border case \(\alpha \leq 1, \, \beta > 1\):
  The mode equals zero, so the interval should be ``attached to the left border'':
  \(\operatorname{BetaHDI}(\alpha, \beta, D) = [0; D]\).
\item
  Right border case \(\alpha > 1, \, \beta \leq 1\):
  The mode equals one, so the interval should be ``attached to the right border'':
  \(\operatorname{BetaHDI}(\alpha, \beta, D) = [1 - D; 1]\)
\end{itemize}

The fourth case is the middle case (\(\alpha > 1,\, \beta > 1\)),
the HDI should be inside \((0;1)\).
Since the density function of the beta distribution is a unimodal function, it consists of two segments:
a monotonically increasing segment \([0, M]\) and
a monotonically decreasing segment \([M, 1]\).
The HDI \([L;R]\) should contain the mode, so

\[
L \in [0; M], \quad
R \in [M; 1].
\]

Since \(R - L = D\), we could also conclude that

\[
L = R - D \in [M - D; 1 - D], \quad
R = L + D \in [D; M + D].
\]

Thus,

\[
L \in [\max(0, M - D);\; \min(M, 1 - D)], \quad
R \in [\max(M, D);\; \min(1, M + D)].
\]

The density function of the beta distribution is also known (see \autocite{sematech}):

\[
f(x) = \dfrac{x^{\alpha - 1} (1 - x)^{\beta - 1}}{\textrm{B}(\alpha, \beta)}, \quad
\textrm{B}(\alpha, \beta) = \dfrac{\Gamma(\alpha)\Gamma(\beta)}{\Gamma(\alpha + \beta)}.
\]

It's easy to see that for the highest density interval \([L; R]\), the following condition is true:

\[
f(L) = f(R).
\]

The left border \(L\) of this interval could be found as a solution of the following equation:

\[
f(t) = f(t + D), \quad \textrm{where }\, t \in [\max(0, M - D);\; \min(M, 1 - D)].
\]

The left side of the equation is monotonically increasing, the right side is monotonically decreasing.
The equation has exactly one solution which could be easily found numerically using the binary search algorithm.

\clearpage

\hypertarget{simulation-study}{%
\section{Simulation study}\label{simulation-study}}

Let's perform a few numerical simulations and see how \(Q_{\operatorname{THD}}\) works in action.\footnote{The source code of all simulations is available on GitHub: \url{https://github.com/AndreyAkinshin/paper-thdqe}}

\hypertarget{simulation-1}{%
\subsection{Simulation 1}\label{simulation-1}}

Let's explore the distribution of estimations for
\(Q_{\operatorname{HF7}}\), \(Q_{\operatorname{HD}}\), and \(Q_{\operatorname{THD-SQRT}}\).
We consider a contaminated normal distribution which is a mixture of two normal distributions:
\((1 - \varepsilon)\mathcal{N}(0, \sigma^2) + \varepsilon\mathcal{N}(0, c\sigma^2)\).
For our simulation, we use \(\varepsilon = 0.01,\; \sigma = 1,\; c = 1\,000\,000\).
We generate \(10\,000\) samples of size 7 randomly taken from the considered distribution.
For each sample, we estimate the median using
\(Q_{\operatorname{HF7}}\), \(Q_{\operatorname{HD}}\), and \(Q_{\operatorname{THD-SQRT}}\).
Thus, we have \(10\,000\) median estimations for each estimator.
Next, we evaluate lower and higher percentiles for each group of estimations.
The results are presented in Table \ref{tab:t2}.

\begin{longtable}[]{@{}rrrr@{}}
\caption{\label{tab:t2} Distribution of median estimations for a contaminated normal distribution}\tabularnewline
\toprule()
quantile & HF7 & HD & THD-SQRT \\
\midrule()
\endfirsthead
\toprule()
quantile & HF7 & HD & THD-SQRT \\
\midrule()
\endhead
0.00 & -1.6921648 & -87.6286082 & -1.6041220 \\
0.01 & -1.1054591 & -9.8771723 & -1.0261234 \\
0.02 & -0.9832125 & -5.2690083 & -0.9067884 \\
0.03 & -0.9037046 & -1.7742334 & -0.8298706 \\
0.04 & -0.8346268 & -0.9921591 & -0.7586603 \\
0.96 & 0.8172518 & 0.8964743 & 0.7540437 \\
0.97 & 0.8789283 & 1.1240294 & 0.8052421 \\
0.98 & 0.9518048 & 4.3675475 & 0.8824462 \\
0.99 & 1.0806293 & 10.4132583 & 0.9900912 \\
1.00 & 2.0596785 & 140.5802861 & 1.7060750 \\
\bottomrule()
\end{longtable}

We also perform the same experiment using the Fréchet distribution (shape=1)
as an example of a right-skewed heavy-tailed distribution.
The results are presented in Table \ref{tab:t3}.

\begin{longtable}[]{@{}rrrr@{}}
\caption{\label{tab:t3} Distribution of median estimations for the Fréchet distribution (shape=1)}\tabularnewline
\toprule()
quantile & HF7 & HD & THD-SQRT \\
\midrule()
\endfirsthead
\toprule()
quantile & HF7 & HD & THD-SQRT \\
\midrule()
\endhead
0.00 & 0.3365648 & 0.4121860 & 0.3720898 \\
0.01 & 0.5161896 & 0.6684699 & 0.5810966 \\
0.02 & 0.5703807 & 0.7578653 & 0.6369594 \\
0.03 & 0.6082605 & 0.8058995 & 0.6834209 \\
0.04 & 0.6433384 & 0.8460783 & 0.7187727 \\
0.96 & 4.2510264 & 7.2021571 & 4.6591661 \\
0.97 & 4.6202217 & 8.3669085 & 5.0186522 \\
0.98 & 5.2815341 & 10.0274664 & 5.6965864 \\
0.99 & 6.5037105 & 14.3159366 & 7.1671722 \\
1.00 & 42.0799646 & \(6\,501.9425729\) & 35.3494053 \\
\bottomrule()
\end{longtable}

In Table \ref{tab:t2}, approximately 2\% of all \(Q_{\operatorname{HD}}\) results exceed 10 by their absolute values
(while the true median value is zero).
Meanwhile, the maximum absolute value of the \(Q_{\operatorname{THD-SQRT}}\) median estimations is approximately \(1.7\).
In Table \ref{tab:t3}, the higher percentiles of \(Q_{\operatorname{HD}}\) estimations are also noticeably higher
than the corresponding values of \(Q_{\operatorname{THD-SQRT}}\).

Thus, \(Q_{\operatorname{THD-SQRT}}\) is much more resistant to outliers than \(Q_{\operatorname{HD}}\).

\clearpage

\hypertarget{simulation-2}{%
\subsection{Simulation 2}\label{simulation-2}}

Let's compare the statistical efficiency of \(Q_{\operatorname{HD}}\) and \(Q_{\operatorname{THD}}\).
We evaluate the relative efficiency of these estimators against \(Q_{\operatorname{HF7}}\)
which is a conventional baseline in such experiments.
For the \(p^\textrm{th}\) quantile, the classic relative efficiency can be calculated
as the ratio of the estimator mean squared errors (\(\operatorname{MSE}\)) (see \autocite{dekking2005}):

\[
\operatorname{Efficiency}(p) =
\dfrac{\operatorname{MSE}(Q_{HF7}, p)}{\operatorname{MSE}(Q_{\textrm{Target}}, p)} =
\dfrac{\operatorname{E}[(Q_{HF7}(p) - \theta(p))^2]}{\operatorname{E}[(Q_{\textrm{Target}}(p) - \theta(p))^2]},
\]

where \(\theta(p)\) is the true quantile value.
We conduct this simulation according to the following scheme:

\begin{itemize}
\tightlist
\item
  We consider a bunch of different symmetric and asymmetric, light-tailed and heavy-tailed distributions
  listed in Table \ref{tab:t4}.
\item
  We enumerate all the percentile values \(p\) from 0.01 to 0.99.
\item
  For each distribution, we generate 200 random samples of the given size.
  For each sample, we estimate the \(p^\textrm{th}\) percentile using
  \(Q_{\operatorname{HF7}}\), \(Q_{\operatorname{HD}}\), and \(Q_{\operatorname{THD-SQRT}}\).
  For each estimator, we calculate the arithmetic average of \((Q(p) - \theta(p))^2\).
\item
  \(\operatorname{MSE}\) is not a robust metric, so we wouldn't get reproducible output in such an experiment.
  To achieve more stable results, we repeat the previous step 101 times and take the median across
  \(\operatorname{E}[(Q(p) - \theta(p))^2]\) values for each estimator.
  This median is our estimation of \(\operatorname{MSE}(Q, p)\).
\item
  We evaluate the relative efficiency of \(Q_{\operatorname{HD}}\) and \(Q_{\operatorname{THD-SQRT}}\)
  against \(Q_{\operatorname{HF7}}\).
\end{itemize}

\begin{longtable}[]{@{}
  >{\raggedright\arraybackslash}p{(\columnwidth - 6\tabcolsep) * \real{0.3875}}
  >{\raggedright\arraybackslash}p{(\columnwidth - 6\tabcolsep) * \real{0.2625}}
  >{\raggedright\arraybackslash}p{(\columnwidth - 6\tabcolsep) * \real{0.1750}}
  >{\raggedright\arraybackslash}p{(\columnwidth - 6\tabcolsep) * \real{0.1750}}@{}}
\caption{\label{tab:t4} Distributions from Simulation 2}\tabularnewline
\toprule()
\begin{minipage}[b]{\linewidth}\raggedright
Distribution
\end{minipage} & \begin{minipage}[b]{\linewidth}\raggedright
Support
\end{minipage} & \begin{minipage}[b]{\linewidth}\raggedright
Skewness
\end{minipage} & \begin{minipage}[b]{\linewidth}\raggedright
Tailness
\end{minipage} \\
\midrule()
\endfirsthead
\toprule()
\begin{minipage}[b]{\linewidth}\raggedright
Distribution
\end{minipage} & \begin{minipage}[b]{\linewidth}\raggedright
Support
\end{minipage} & \begin{minipage}[b]{\linewidth}\raggedright
Skewness
\end{minipage} & \begin{minipage}[b]{\linewidth}\raggedright
Tailness
\end{minipage} \\
\midrule()
\endhead
\texttt{Uniform(a=0,\ b=1)} & \([0;1]\) & Symmetric & Light-tailed \\
\texttt{Triangular(a=0,\ b=2,\ c=1)} & \([0;2]\) & Symmetric & Light-tailed \\
\texttt{Triangular(a=0,\ b=2,\ c=0.2)} & \([0;2]\) & Right-skewed & Light-tailed \\
\texttt{Beta(a=2,\ b=4)} & \([0;1]\) & Right-skewed & Light-tailed \\
\texttt{Beta(a=2,\ b=10)} & \([0;1]\) & Right-skewed & Light-tailed \\
\texttt{Normal(m=0,\ sd=1)} & \((-\infty;+\infty)\) & Symmetric & Light-tailed \\
\texttt{Weibull(scale=1,\ shape=2)} & \([0;+\infty)\) & Right-skewed & Light-tailed \\
\texttt{Student(df=3)} & \((-\infty;+\infty)\) & Symmetric & Light-tailed \\
\texttt{Gumbel(loc=0,\ scale=1)} & \((-\infty;+\infty)\) & Right-skewed & Light-tailed \\
\texttt{Exp(rate=1)} & \([0;+\infty)\) & Right-skewed & Light-tailed \\
\texttt{Cauchy(x0=0,\ gamma=1)} & \((-\infty;+\infty)\) & Symmetric & Heavy-tailed \\
\texttt{Pareto(loc=1,\ shape=0.5)} & \([1;+\infty)\) & Right-skewed & Heavy-tailed \\
\texttt{Pareto(loc=1,\ shape=2)} & \([1;+\infty)\) & Right-skewed & Heavy-tailed \\
\texttt{LogNormal(mlog=0,\ sdlog=1)} & \((0;+\infty)\) & Right-skewed & Heavy-tailed \\
\texttt{LogNormal(mlog=0,\ sdlog=2)} & \((0;+\infty)\) & Right-skewed & Heavy-tailed \\
\texttt{LogNormal(mlog=0,\ sdlog=3)} & \((0;+\infty)\) & Right-skewed & Heavy-tailed \\
\texttt{Weibull(shape=0.3)} & \([0;+\infty)\) & Right-skewed & Heavy-tailed \\
\texttt{Weibull(shape=0.5)} & \([0;+\infty)\) & Right-skewed & Heavy-tailed \\
\texttt{Frechet(shape=1)} & \((0;+\infty)\) & Right-skewed & Heavy-tailed \\
\texttt{Frechet(shape=3)} & \((0;+\infty)\) & Right-skewed & Heavy-tailed \\
\bottomrule()
\end{longtable}

\begin{figure}[ht!]

{\centering \includegraphics{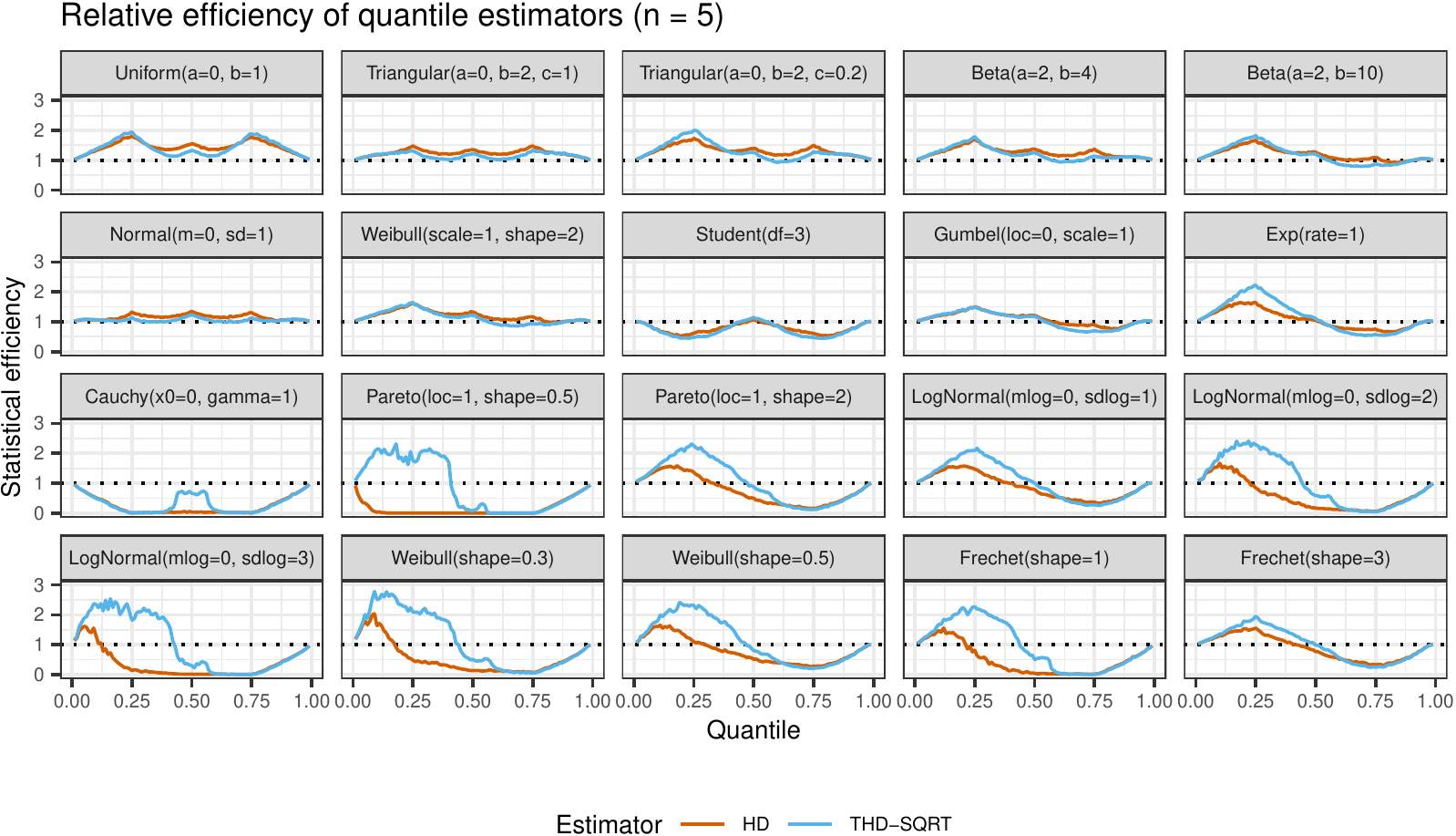} 

}

\caption{Relative statistical efficiency of quantile estimators for n=5}\label{fig:eff5}
\end{figure}

\begin{figure}[ht!]

{\centering \includegraphics{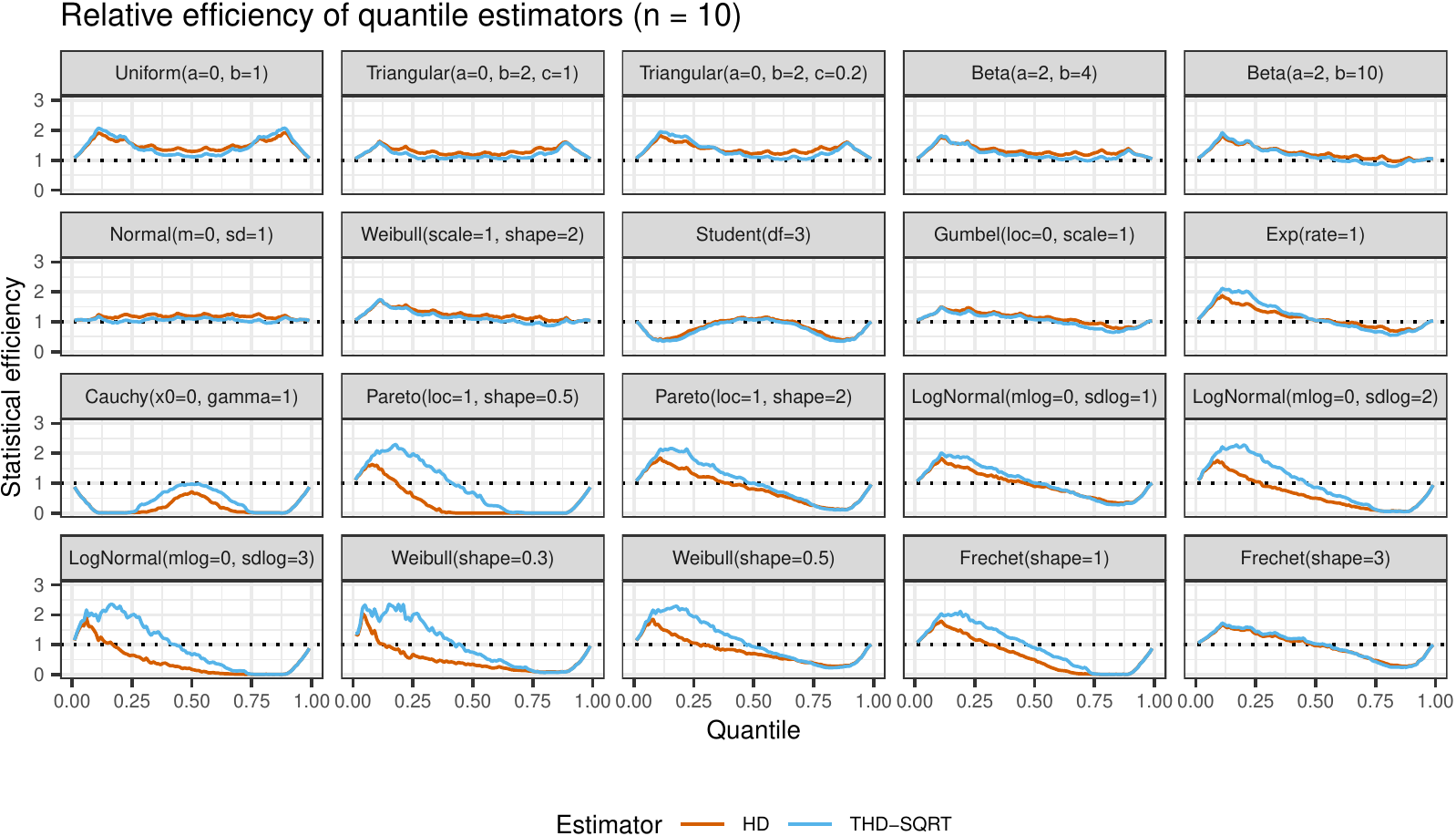} 

}

\caption{Relative statistical efficiency of quantile estimators for n=10}\label{fig:eff10}
\end{figure}

\begin{figure}[ht!]

{\centering \includegraphics{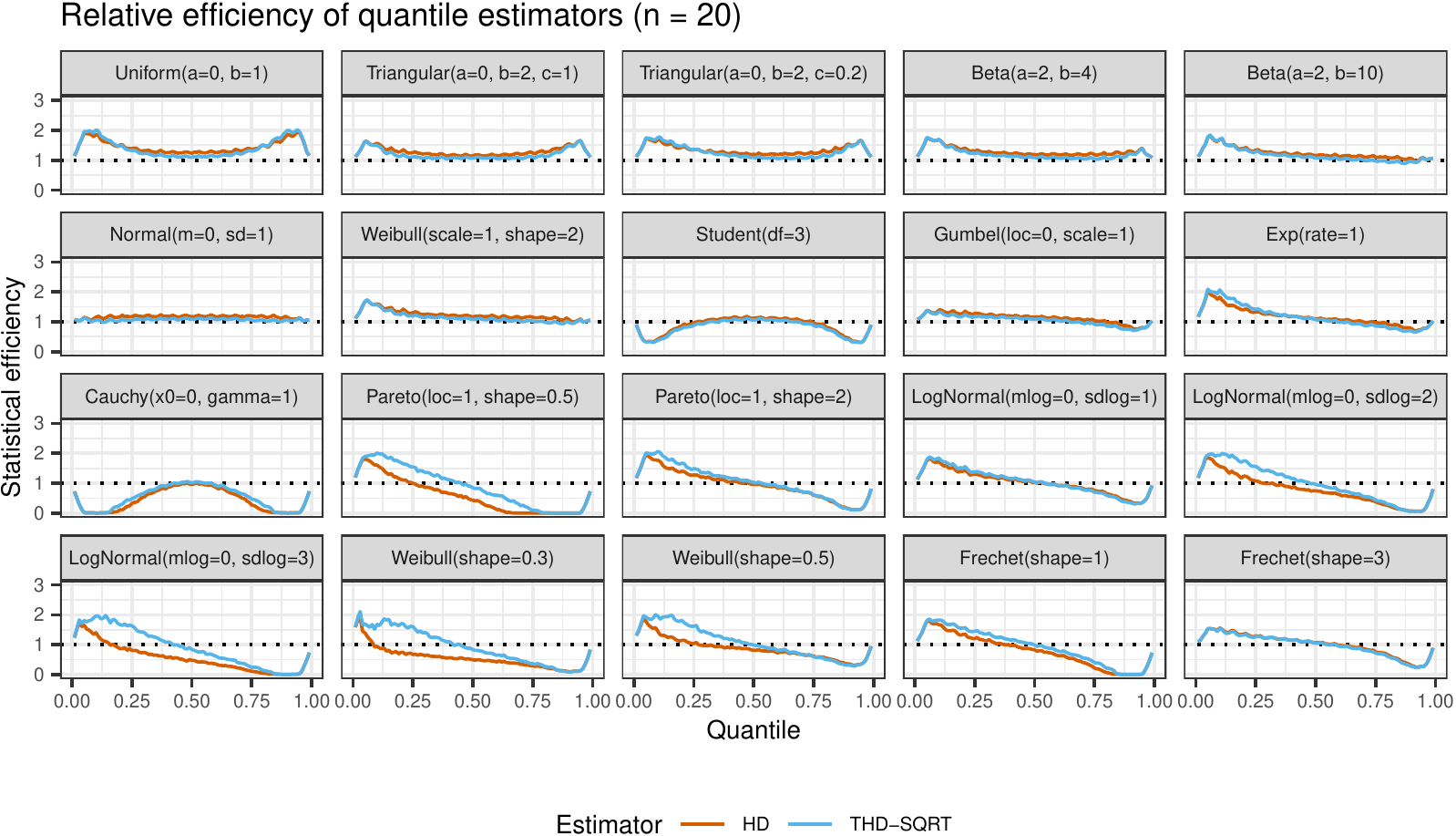} 

}

\caption{Relative statistical efficiency of quantile estimators for n=20}\label{fig:eff20}
\end{figure}

The results of this simulation for \(n = \{ 5, 10, 20 \}\) are presented in
Figure \ref{fig:eff5}, Figure \ref{fig:eff10}, and Figure \ref{fig:eff20}.

\clearpage

\clearpage

As we can see, \(Q_{\operatorname{THD-SQRT}}\) is not so efficient as \(Q_{\operatorname{HD}}\)
in the case of light-tailed distributions.
However, in the case of heavy-tailed distributions,
\(Q_{\operatorname{THD-SQRT}}\) has better efficiency than \(Q_{\operatorname{HD}}\)
because the estimations of \(Q_{\operatorname{HD}}\) are corrupted by outliers.

\hypertarget{conclusion}{%
\section{Conclusion}\label{conclusion}}

There is no perfect quantile estimator that fits all kinds of problems.
The choice of a specific estimator has to be made
based on the knowledge of the domain area and the properties of the target distributions.
\(Q_{\operatorname{HD}}\) is a good alternative to \(Q_{\operatorname{HF7}}\) in the light-tailed distributions
because it has higher statistical efficiency.
However, if extreme outliers may appear, estimations of \(Q_{\operatorname{HD}}\) could be heavily corrupted.
\(Q_{\operatorname{THD}}\) could be used as
a reasonable trade-off between \(Q_{\operatorname{HF7}}\) and \(Q_{\operatorname{HD}}\).
In most cases, \(Q_{\operatorname{THD}}\) has better efficiency than \(Q_{\operatorname{HF7}}\)
and it's also more resistant to outliers than \(Q_{\operatorname{HD}}\).
By customizing the width \(D\) of the highest density interval, we could set the desired breakdown point
according to the research goals.
Also, \(Q_{\operatorname{THD}}\) has better computational efficiency than \(Q_{\operatorname{HD}}\)
which makes it a faster option in practical applications.

\hypertarget{disclosure-statement}{%
\section{Disclosure statement}\label{disclosure-statement}}

The author reports there are no competing interests to declare.

\hypertarget{acknowledgments}{%
\section{Acknowledgments}\label{acknowledgments}}

The author thanks Ivan Pashchenko for valuable discussions.

\newpage

\hypertarget{reference-implementation}{%
\section{Reference implementation}\label{reference-implementation}}

Here is an R implementation of the suggested estimator:

\begin{Shaded}
\begin{Highlighting}[]
\NormalTok{getBetaHdi }\OtherTok{\textless{}{-}} \ControlFlowTok{function}\NormalTok{(a, b, width) \{}
\NormalTok{  eps }\OtherTok{\textless{}{-}} \FloatTok{1e{-}9}
  \ControlFlowTok{if}\NormalTok{ (a }\SpecialCharTok{\textless{}} \DecValTok{1} \SpecialCharTok{+}\NormalTok{ eps }\SpecialCharTok{\&}\NormalTok{ b }\SpecialCharTok{\textless{}} \DecValTok{1} \SpecialCharTok{+}\NormalTok{ eps) }\CommentTok{\# Degenerate case}
    \FunctionTok{return}\NormalTok{(}\FunctionTok{c}\NormalTok{(}\ConstantTok{NA}\NormalTok{, }\ConstantTok{NA}\NormalTok{))}
  \ControlFlowTok{if}\NormalTok{ (a }\SpecialCharTok{\textless{}} \DecValTok{1} \SpecialCharTok{+}\NormalTok{ eps }\SpecialCharTok{\&}\NormalTok{ b }\SpecialCharTok{\textgreater{}} \DecValTok{1}\NormalTok{) }\CommentTok{\# Left border case}
    \FunctionTok{return}\NormalTok{(}\FunctionTok{c}\NormalTok{(}\DecValTok{0}\NormalTok{, width))}
  \ControlFlowTok{if}\NormalTok{ (a }\SpecialCharTok{\textgreater{}} \DecValTok{1} \SpecialCharTok{\&}\NormalTok{ b }\SpecialCharTok{\textless{}} \DecValTok{1} \SpecialCharTok{+}\NormalTok{ eps) }\CommentTok{\# Right border case}
    \FunctionTok{return}\NormalTok{(}\FunctionTok{c}\NormalTok{(}\DecValTok{1} \SpecialCharTok{{-}}\NormalTok{ width, }\DecValTok{1}\NormalTok{))}
  \ControlFlowTok{if}\NormalTok{ (width }\SpecialCharTok{\textgreater{}} \DecValTok{1} \SpecialCharTok{{-}}\NormalTok{ eps)}
    \FunctionTok{return}\NormalTok{(}\FunctionTok{c}\NormalTok{(}\DecValTok{0}\NormalTok{, }\DecValTok{1}\NormalTok{))}
  
  \CommentTok{\# Middle case}
\NormalTok{  mode }\OtherTok{\textless{}{-}}\NormalTok{ (a }\SpecialCharTok{{-}} \DecValTok{1}\NormalTok{) }\SpecialCharTok{/}\NormalTok{ (a }\SpecialCharTok{+}\NormalTok{ b }\SpecialCharTok{{-}} \DecValTok{2}\NormalTok{)}
\NormalTok{  pdf }\OtherTok{\textless{}{-}} \ControlFlowTok{function}\NormalTok{(x) }\FunctionTok{dbeta}\NormalTok{(x, a, b)}
\NormalTok{  l }\OtherTok{\textless{}{-}} \FunctionTok{uniroot}\NormalTok{(}
    \AttributeTok{f =} \ControlFlowTok{function}\NormalTok{(x) }\FunctionTok{pdf}\NormalTok{(x) }\SpecialCharTok{{-}} \FunctionTok{pdf}\NormalTok{(x }\SpecialCharTok{+}\NormalTok{ width),}
    \AttributeTok{lower =} \FunctionTok{max}\NormalTok{(}\DecValTok{0}\NormalTok{, mode }\SpecialCharTok{{-}}\NormalTok{ width),}
    \AttributeTok{upper =} \FunctionTok{min}\NormalTok{(mode, }\DecValTok{1} \SpecialCharTok{{-}}\NormalTok{ width),}
    \AttributeTok{tol =} \FloatTok{1e{-}9}
\NormalTok{  )}\SpecialCharTok{$}\NormalTok{root}
\NormalTok{  r }\OtherTok{\textless{}{-}}\NormalTok{ l }\SpecialCharTok{+}\NormalTok{ width}
  \FunctionTok{return}\NormalTok{(}\FunctionTok{c}\NormalTok{(l, r))}
\NormalTok{\}}

\NormalTok{thdquantile }\OtherTok{\textless{}{-}} \ControlFlowTok{function}\NormalTok{(x, probs, }\AttributeTok{width =} \DecValTok{1} \SpecialCharTok{/} \FunctionTok{sqrt}\NormalTok{(}\FunctionTok{length}\NormalTok{(x)))}
               \FunctionTok{sapply}\NormalTok{(probs, }\ControlFlowTok{function}\NormalTok{(p) \{}
\NormalTok{  n }\OtherTok{\textless{}{-}} \FunctionTok{length}\NormalTok{(x)}
  \ControlFlowTok{if}\NormalTok{ (n }\SpecialCharTok{==} \DecValTok{0}\NormalTok{) }\FunctionTok{return}\NormalTok{(}\ConstantTok{NA}\NormalTok{)}
  \ControlFlowTok{if}\NormalTok{ (n }\SpecialCharTok{==} \DecValTok{1}\NormalTok{) }\FunctionTok{return}\NormalTok{(x)}

\NormalTok{  x }\OtherTok{\textless{}{-}} \FunctionTok{sort}\NormalTok{(x)}
\NormalTok{  a }\OtherTok{\textless{}{-}}\NormalTok{ (n }\SpecialCharTok{+} \DecValTok{1}\NormalTok{) }\SpecialCharTok{*}\NormalTok{ p}
\NormalTok{  b }\OtherTok{\textless{}{-}}\NormalTok{ (n }\SpecialCharTok{+} \DecValTok{1}\NormalTok{) }\SpecialCharTok{*}\NormalTok{ (}\DecValTok{1} \SpecialCharTok{{-}}\NormalTok{ p)}
\NormalTok{  hdi }\OtherTok{\textless{}{-}} \FunctionTok{getBetaHdi}\NormalTok{(a, b, width)}
\NormalTok{  hdiCdf }\OtherTok{\textless{}{-}} \FunctionTok{pbeta}\NormalTok{(hdi, a, b)}
\NormalTok{  cdf }\OtherTok{\textless{}{-}} \ControlFlowTok{function}\NormalTok{(xs) \{}
\NormalTok{    xs[xs }\SpecialCharTok{\textless{}=}\NormalTok{ hdi[}\DecValTok{1}\NormalTok{]] }\OtherTok{\textless{}{-}}\NormalTok{ hdi[}\DecValTok{1}\NormalTok{]}
\NormalTok{    xs[xs }\SpecialCharTok{\textgreater{}=}\NormalTok{ hdi[}\DecValTok{2}\NormalTok{]] }\OtherTok{\textless{}{-}}\NormalTok{ hdi[}\DecValTok{2}\NormalTok{]}
\NormalTok{    (}\FunctionTok{pbeta}\NormalTok{(xs, a, b) }\SpecialCharTok{{-}}\NormalTok{ hdiCdf[}\DecValTok{1}\NormalTok{]) }\SpecialCharTok{/}\NormalTok{ (hdiCdf[}\DecValTok{2}\NormalTok{] }\SpecialCharTok{{-}}\NormalTok{ hdiCdf[}\DecValTok{1}\NormalTok{])}
\NormalTok{  \}}
\NormalTok{  iL }\OtherTok{\textless{}{-}} \FunctionTok{floor}\NormalTok{(hdi[}\DecValTok{1}\NormalTok{] }\SpecialCharTok{*}\NormalTok{ n)}
\NormalTok{  iR }\OtherTok{\textless{}{-}} \FunctionTok{ceiling}\NormalTok{(hdi[}\DecValTok{2}\NormalTok{] }\SpecialCharTok{*}\NormalTok{ n)}

\NormalTok{  cdfs }\OtherTok{\textless{}{-}} \FunctionTok{cdf}\NormalTok{(iL}\SpecialCharTok{:}\NormalTok{iR}\SpecialCharTok{/}\NormalTok{n)}
\NormalTok{  W }\OtherTok{\textless{}{-}} \FunctionTok{tail}\NormalTok{(cdfs, }\SpecialCharTok{{-}}\DecValTok{1}\NormalTok{) }\SpecialCharTok{{-}} \FunctionTok{head}\NormalTok{(cdfs, }\SpecialCharTok{{-}}\DecValTok{1}\NormalTok{)}
  \FunctionTok{sum}\NormalTok{(x[(iL}\SpecialCharTok{+}\DecValTok{1}\NormalTok{)}\SpecialCharTok{:}\NormalTok{iR] }\SpecialCharTok{*}\NormalTok{ W)}
\NormalTok{\})}
\end{Highlighting}
\end{Shaded}

An implementation of the original Harrell-Davis quantile estimator could be found in the \texttt{Hmisc} package (see \autocite{hmisc}).

\newpage

\printbibliography

\end{document}